# Single-cell eQTLGen Consortium: a personalized understanding of disease

Van der Wijst MG[1], de Vries DH[1], Groot HE[2], Trynka G[3], Hon CC[4], Nawijn MC[5], Idaghdour Y[6], van der Harst P[2], Ye CJ[7], Powell J[8], Theis FJ[9], Mahfouz A[10], Heinig M[11], Franke L[1]

[1] Department of Genetics, Oncode Institute, University of Groningen, University Medical Center Groningen, Groningen, The Netherlands

[2] Department of Cardiology, University of Groningen, University Medical Center Groningen, Groningen, The Netherlands

[3] Wellcome Trust Sanger Institute, Open Targets, Wellcome Genome Campus, Hinxton, Cambridge, UK

[4] RIKEN Center for Integrative Medical Sciences, Yokohama, Japan

[5] Department of Pathology and Medical Biology, GRIAC Research institute, University of Groningen, University Medical Center Groningen, the Netherlands

[6] Program in Biology, Public Health Research Center, New York University Abu Dhabi, UAE

[7] Institute for Human Genetics, Bakar Computational Health Sciences Institute, Bakar ImmunoX Initiative, Division of Rheumatology, Department of Medicine, Department of Bioengineering and Therapeutic Sciences, Department of Epidemiology and Biostatistics, Chan Zuckerberg Biohub, University of California San Francisco, San Francisco, USA.

[8] Garvan-Weizmann Centre for Cellular Genomics, Garvan Institute, Sydney, Australia
UNSW Cellular Genomics Futures Institute, University of New South Wales, Sydney, Australia

[9] Institute of Computational Biology, Helmholtz Zentrum München, Neuherberg, Germany
Department of Mathematics, Technical University of Munich, Garching bei München, Germany

[10] Leiden Computational Biology Center, Leiden University Medical Center, Leiden, The Netherlands
Delft Bioinformatics Lab, Delft University of Technology, Delft, The Netherlands

[11] Institute of Computational Biology, Helmholtz Zentrum München, Neuherberg, Germany
Department of Informatics, Technical University of Munich, Garching bei München, Germany

**Abstract**

In recent years, functional genomics approaches combining genetic information with bulk RNA-sequencing data have identified the downstream expression effects of disease-associated genetic risk factors through so-called expression quantitative trait locus (eQTL) analysis. Single-cell RNA-sequencing creates enormous opportunities for mapping eQTLs across different cell types and in dynamic processes, many of which are obscured when using bulk methods. The enormous increase in throughput and reduction in cost per cell now allow this technology to be applied to large-scale population genetics studies. Therefore, we have founded the single-cell eQTLGen consortium (sc-eQTLGen), aimed at pinpointing disease-causing genetic variants and identifying the cellular contexts in which they affect gene expression. Ultimately, this information can enable development of personalized medicine. Here, we outline the goals, approach, potential utility and early proofs-of-concept of the sc-eQTLGen consortium. We also provide a set of study design considerations for future single-cell eQTL studies.

**Interindividual variation needs to be studied at the single-cell level**

Genetic variants, most frequently single nucleotide polymorphisms (SNPs), can contribute to disease in a plethora of ways. In monogenic diseases, one single variant is sufficient to result in a disease phenotype. In most complex diseases, tens to hundreds of variants each independently contribute to disease risk and an accumulation of risk alleles – often in combination with specific environmental exposures – is required to develop the disease phenotype. The overwhelming evidence showing enrichment of disease-associated variants in regulatory regions suggests that regulation of gene expression is likely a dominant mediator for disease risk. Expression quantitative trait loci (eQTL) analysis is commonly used for linking disease risk-SNPs to downstream expression effects on local (*cis*) or distal (*trans*) genes. Large-scale eQTL efforts such as GTEx[1], PsychENCODE[2], ImmVar[3], BLUEPRINT[4], CAGE[5], and eQTLGen[6] have proven highly valuable to identify downstream transcriptional consequences. All these efforts together lead to ever growing sample sizes that now allow us to start identifying both *cis*- and *trans*-eQTLs.

An important next step is to define the contexts in which disease risk-SNPs affect gene expression levels. This will help to better understand the molecular and cellular mechanisms by which disease risk is conferred and to inform therapeutic strategies. This is particularly important, as recent analyses have shown that many eQTL effects are tissue-[1,7] and cell type-specific[8,9]. Additionally, many eQTLs are conditional, and only revealed after specific stimuli that, for example, change the activation or differentiation of specific cell types[3,10]. Beyond the ability to annotate individual disease associations, cell-type specific eQTLs have been shown to be strongly enriched for heritability across complex traits[11]. Sorting[9,12] and computational deconvolution[13,14] of cell types from bulk samples have been used to uncover context-specificity of eQTLs. However, these methods are biased towards known cell types defined by a limited set of marker genes[15], are of limited use for less abundant cell types, and do not capture any heterogeneity within a sorted population. In contrast, single-cell RNA-sequencing (scRNA-seq) enables the simultaneous and unbiased estimation of cellular composition and cell type-specific gene expression[16], and is particularly well positioned to investigate rare cell types[17]. As opposed to using bulk data, single-cell data allows us to also link genetics to phenomena such as cell-to-cell expression variability[10], cell type heterogeneity[18], and gene regulatory network differences[16]. As such, single-cell analyses in a population-based setting will likely become mainstream in the next few years. However, we envision that most scientific value will be obtained by unifying these efforts. Additionally, to utilize the aforementioned developments in the single-cell field most efficiently and effectively, a coordinated effort from multiple research groups is urgently needed.

Here we introduce the single-cell eQTLGen consortium (sc-eQTLGen), a large-scale, international collaborative effort that has been set up to identify the upstream interactors and downstream consequences of disease-related genetic variants in specific immune cell types (https://eqtlgen.org/single-cell.html, Figure 1). In this consortium we will attain a sufficiently large sample size to have the statistical power to unbiasedly identify cell type-specific effects on both local (*cis*) and distal (*trans*) genes. Moreover, we aim to reconstruct context-specific gene regulatory networks (GRNs) by combining single-cell and bulk RNA-seq datasets to achieve optimal resolution. We expect a broad impact of the results of sc-eQTLGen that ranges from prioritizing disease-risk genes to predicting drug efficacy through the reconstruction of personalized GRNs.

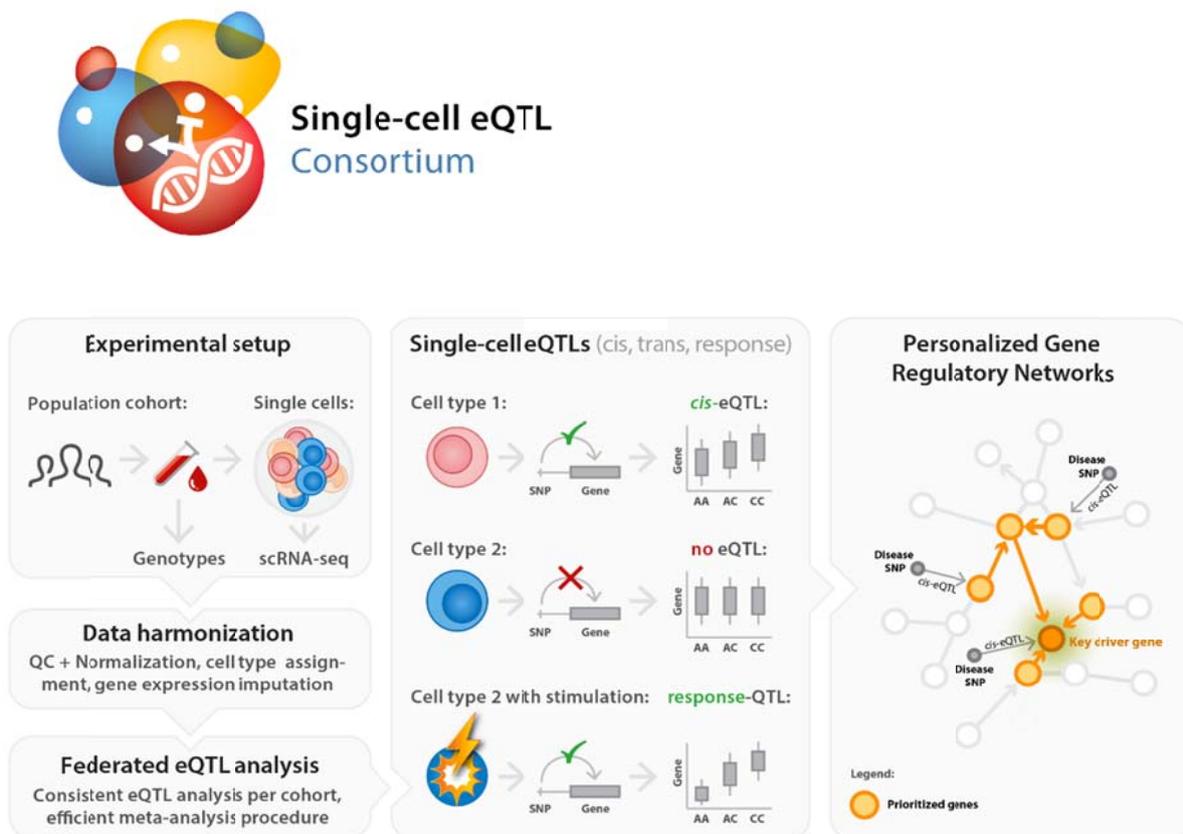

**Figure 1. Set-up of the single-cell eQTLGen (sc-eQTLGen) consortium.** The sc-eQTLGen consortium combines an individual's genetic information with single-cell RNA expression (scRNA-seq) data of peripheral blood mononuclear cells (PBMCs) in order to identify effects of genetic variation on downstream gene expression levels (eQTLs) and to enable reconstruction of personalized gene regulatory networks.

**Integration of sc-eQTLGen within the scientific landscape**

Large numbers of single cell expression profiles from many individuals are required to reach our goals. The accessibility and clinical relevance of peripheral blood mononuclear cells (PBMCs) have made them the most studied cell types in current population-based scRNA-seq datasets. Therefore, to have such datasets from the same tissue type readily available, we have chosen to focus on PBMCs. It also allows for continuation of the knowledge acquired through the eQTLGen consortium, which performed the largest eQTL meta-analysis to date using whole blood bulk gene expression data of over 30,000 individuals to reveal the influence of genetics on gene expression[6]. The sc-eQTLGen consortium now allows us to take the next step and determine the cell types and contexts in which the eQTL effects manifest. Beyond resolving the influence of genetics on individual genes, the consortium will also take advantage of the unique features of scRNA-seq data to learn the directionality of GRNs and uncover how genetics is affecting co-expression relationships[16]. We expect that the infrastructure and best practices developed within sc-eQTLGen can serve as a basis for studying population genetics at the single-cell level in solid tissues in the future.

Other large-scale efforts such as the Human Cell Atlas (HCA)[19] or Lifetime FET flagship consortium (https://lifetime-fetflagship.eu) mainly focus on mapping all cells of the human body or a disease context in a limited number of individuals. The sc-eQTLGen consortium is an important addition to those efforts by putting a unique focus on deciphering the impact of genetic variation on gene expression and its regulation. To achieve our goals, we require a large number of individuals while having a relatively smaller number of cells per individual. This enables accurately capturing both the genetic variation and cell type heterogeneity. By building on the data and harmonized cell type annotations generated within the HCA, our results will be easily transferable to other datasets as well. We will share best practices of the HCA consortium with regard to data acquisition, analysis and reporting. We also share standards for open science and the infrastructure and legal frameworks for data sharing while accounting for the privacy issues specific to genetic, health record and demographic information.

**Single-cell eQTL analysis: the new era of population genetics**

The practice of identifying eQTLs is shifting from bulk to single-cell analyses. Considering only its ability to identify eQTLs, scRNA-seq data has a lower statistical power compared to equal-sized bulk RNA-seq data, likely due to increased sparsity of the single-cell data[20]. Nevertheless, there are several clear benefits of single-cell over bulk expression data for QTL analysis. First, scRNA-seq data enables the simultaneous estimation of the composition and expression profiles of discrete cell populations including cell types and their activation states[16] (Figure 2). Second, scRNA-seq data provides a flexible, unbiased approach that has increased resolution to define cell states along continuous dynamic processes in which the eQTL effects manifest themselves[10]. Third, single-cell data allows estimating the variability in gene expression across individual cells[21,22], which could be used as a parameter in a linear mixed model to obtain better estimates of the mean while accounting for differences in the degree of heterogeneity between cell types. Fourth, the large number of observations per individual (i.e. cells) enable the generation of personalized co-expression networks, which vastly reduces the number of individuals required to identify SNPs altering co-expression relationships (i.e. co-expression QTLs[16]). Fifth, the single-cell nature now allows us to look at the effect of genetic variation on transcriptomic traits other than average gene expression level, such as dispersion QTLs that alter the variance independently of the mean expression[20] or cell type proportion QTLs[23], providing a new angle on how genetic variation may impact disease pathogenesis. Finally, and paradoxically, is the potential benefit of lower experimental costs compared to bulk experiments on sorted cells: such experiments require a library to be generated for each sorted population, whereas a single scRNA-seq library of just one sample contains all this information and can easily be multiplexed across multiple individuals[23].

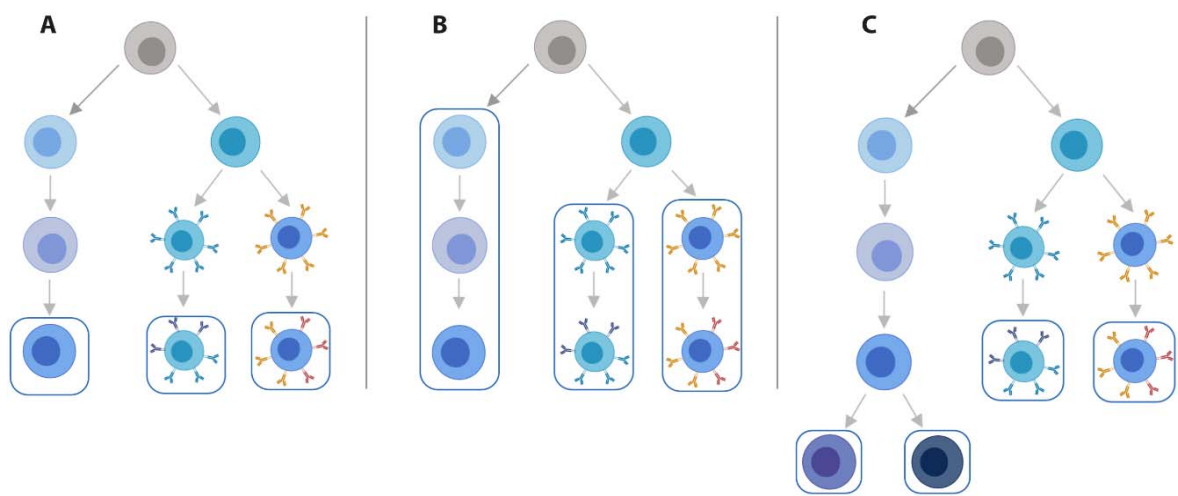

**Figure 2. scRNA-seq data offers increased flexibility in the eQTL analysis strategy over bulk RNA-seq data.** Using scRNA-seq data for eQTL mapping offers numerous advantages over bulk RNA-seq based approaches, of which the flexibility in analysis strategy is a major one. **(A)** From single cell data, individual cell types can be identified and we can map eQTLs for each of these. **(B)** Alternatively, lineages based on either knowledge of cell developmental lineages or through pseudo-time based approaches can be constructed. By positioning cells across a trajectory dynamic changes in the allelic effects on gene expression levels as a function of trajectory position can be integrated. **(C)** Finally, as the discoveries of new cell subtypes are made or cell type definitions are being refined, the analysis can be revisiting by re-classifying cells and determining how the genetic effects on gene expression vary on these new annotations.

So far, only a limited number of papers have performed eQTL analysis using scRNA-seq data[10,16,20,23]. In the earliest single-cell eQTL studies, bulk-based eQTL analysis approaches, such as Spearman rank correlation[24,25] and linear regression[26,27], were applied to the average expression level of all cells from a particular cell type per individual. However, the underlying assumptions of these bulk-based approaches may not be applicable to scRNA-seq data. Therefore, these bulk-based methods will lose statistical power when applied to scRNA-seq data, because of the inflation of zero values (i.e. sparsity). More recently, single-cell-specific eQTL methods have been developed that, for example, take into account zero-inflated gene expression[20,28] or take advantage of pseudotime (i.e. statistically inferred time from snapshot data) to increase the resolution by which response-/differentiation-associated eQTLs (dynamic eQTLs, i.e. eQTLs that dynamically change along pseudotime) can be identified[10]. Instead of averaging gene expression levels across all cells from a particular cell type, some of these approaches look at the fraction of zero expression and the non-zero expression separately for each gene[28]. Other approaches take dynamic pseudotime-defined instead of statically-defined cell types into consideration for the eQTL analysis[10]. This latter approach was shown to uncover hundreds of new eQTL variants during iPSC differentiation that had not been detected before using bulk analysis[10]. In line with this, we expect that some of these methodological advances, as opposed to bulk-based approaches, will further improve the power and resolution of single-cell eQTL analysis. However, there are two initial challenges that need to be carefully addressed for single-cell eQTL mapping: firstly, the normalization of data to remove technical variation in sequencing depth per cell, while avoiding the removal of biological variation; and secondly, the identification or classification of a cell into a cell type or state.

During library preparation and sequencing, technical and stochastic factors will lead to variation in cell-to-cell sequencing depth. However, simply normalizing to equal sequencing depth per cell will remove important biological variation – for example a $CD4^+$ T cell is expected to have lower RNA contents than a plasma B cell. Therefore, we need to employ normalization strategies that can account for traditional batch effects, such as sample run or sequencing lane, while retaining genetically-driven differences and adjusting for technical cell-to-cell variation for very large numbers of cells[29,30].

Once normalized, each cell needs to be accurately annotated into a cell type or cell state to maximize the statistical power to detect cell type-specific eQTLs. We encourage the use of individual cell classification approaches, rather than cluster-labelling methods. Clustering approaches are powerful ways of identifying a subpopulation of cells that share similar expression levels. However, while most cells placed in a specific cluster will likely be the same cell type, clusters can also contain alternative cell types. Labelling all cells in a cluster based on a high percentage of the expression of a canonical marker(s) will therefore lead to the incorrect classification of some cells[31]. To acquire a reliable classification model, large scRNA-seq datasets from various contexts are required. Such datasets have been collected within large-scale efforts such as our consortium and the HCA. We expect these will help to develop a gold standard classification model that can classify each cell independently. This will ensure a higher accuracy in cell labelling and thus will maximize power to detect cell type-specific effects.

After solving these challenges, eQTLs can be mapped by either averaging the normalized expression levels on a per gene, per cell type, per individual basis. Alternatively, each cell from an individual can be taken as a repeated measure which can then be used to fit a statistical model to all cells, while including a random effect of the individual.

Instead of using observational studies, eQTLs could also be identified through experimental approaches that use single cells as individual units of experimentation[32]. Sample multiplexing (Box 1) can be combined with experimental perturbation to more efficiently characterize the genetic architecture of gene expression. For example, synthetic genetic perturbations with CRISPR/Cas9 may allow precise control of the expression levels of target gene regulators enabling the validation of detected *trans*-eQTLs and the establishment of upper and lower bounds of *trans* effects. Encoding environmental and genetic perturbations across large population cohorts also enables new designs for studying genetic interactions, both gene-by-environment and gene-by-gene (epistasis). Historically, characterizing these effects in human cells has been plagued by the lack of power and the susceptibility to technical confounding of bulk experiments. Recent work that knocked out ~150 regulators in primary human T cells of nine donors illustrates a proof of concept of how single-cell sequencing across individuals can be combined with experimental perturbations to detect these genetic interactions[33].

**Single-cell GRN reconstruction: taking eQTLs one step further**

In the case of complex diseases, it is not the disruption of a single gene that causes the disease phenotype. In fact, hundreds of variants can contribute to the disease and converge into just a few key disrupted regulatory pathways[34,35]. Therefore, for a better disease understanding and to take eQTLs one step further, one has to look beyond the disruption of individual genes and determine how the interaction of genes changes based on cell type[13,36,37], environment[38,39] and an individual's genetic makeup[15,16]. The sc-eQTLGen consortium will do so by reconstructing personalized, cell type-specific GRNs[40] (Figure 3). The unique features of scRNA-seq data, among which the inference of pseudotime and RNA velocity[41], enable learning the directionality of network connections[42]. We expect that such personalized GRNs will help explain for example differences in interindividual drug responses, and thereby, will aid in precision medicine in the future.

Reconstruction of GRNs from single cell data (reviewed in [43]) is complicated by the sparsity of the data as a consequence of the stochasticity underlying gene expression[44] and dropouts, i.e. genes that are not detected in some cells as a consequence of technical limitations[45]. This sparsity leads to lower correlation estimates that obscure the identification of true edges in the GRNs. Several solutions have been developed to overcome this problem, including the implementation of prior information[46,47], gene expression imputation[46,48] and usage of alternative measurements of correlation[49,50].

Firstly, prior information encoded in the DNA sequence can be used to overcome these complications[51,52]. Such priors on regulatory interactions can be derived from, for example, ChIP-seq data[53], ATAC-seq data[54] or from perturbation experiments[33,46]. Implementation of such priors was shown to improve bulk GRN reconstruction[54-56], and similarly, it is expected to also improve GRNs reconstructed from single-cell data[46,47]. However, caution is warranted when using this information, as their effect on GRN reconstruction depends on the quality of these data priors[57,58] and priors derived from bulk data may not hold true at the single-cell level[59]. Recent technological advances enable studying chromatin accessibility[60, 61] and expression of enhancers RNAs[62, 63] at the single cell level, which will make it possible to implement single-cell derived priors in GRN reconstruction in the future.

Secondly, gene expression imputation may be used to restore the underlying correlation structure. However, current gene expression imputation methods become more unreliable as the dropout rates increase[48,49]. After gene expression imputation, more network edges are identified, but with a higher chance of detecting false positives[46,48]. Nevertheless, by combining prior information with imputation, GRN reconstruction can be improved both in the bulk[54] and single cell setting[46]. For example, one can replace transcription factor expression with inferred transcription factor activities based on the collective expression patterns of their target genes or take advantage of cross-omics relationships[64].

Finally, alternative correlation measures are being explored to overcome the complications associated with data sparsity, including measures of proportionality[50] and by calculating the correlations on measures other than the normalized expression counts[49]. For example, Z-scores of the gene expression distributions of highly similar cells have been used to calculate the co-expression relationships. This approach could reveal the true correlation structure that was otherwise hidden by technical artifacts[49]. In addition to these computational tools, technological advances, such as single-cell multi-omics approaches[65,66] and improved experimental protocols, are expected to alleviate these complications. Moreover, being able to assess multiple layers of information within the same cell, e.g. chromatin accessibility, DNA methylation, gene and protein expression, opens unique opportunities for developing new methodology for GRN reconstruction and validation. Altogether, this will further improve the accuracy of GRNs reconstructed using single-cell data in the future.

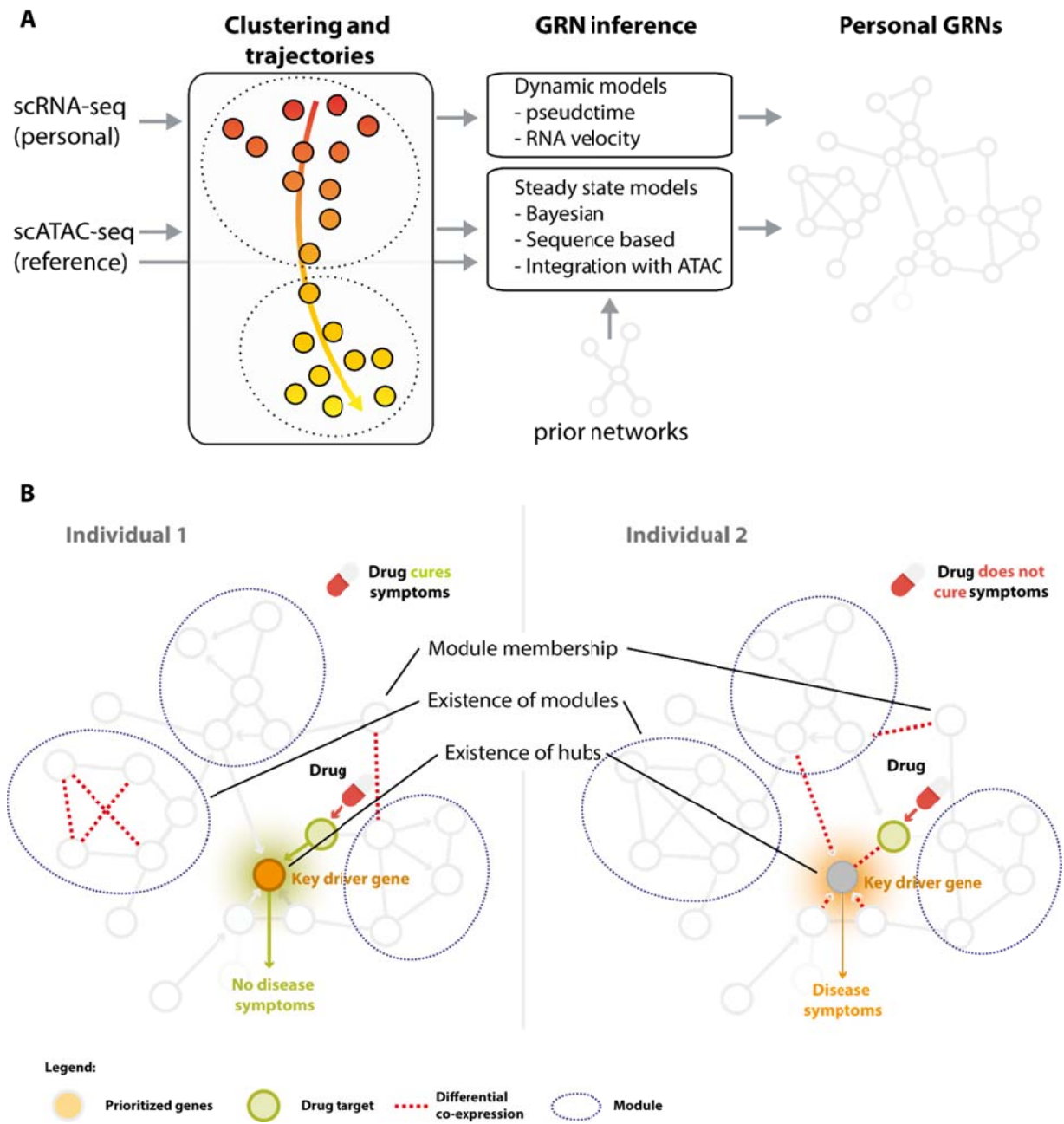

**Figure 3. Reconstruction of personalized gene regulatory networks. A)** Individual and cell-type specific scRNA-seq data will be used to construct personalized gene regulatory networks. Some single cell datasets allow for the inference of trajectories, for instance in response to a stimulus. These can be used as input to dynamic models to infer causal (directed) interactions. Steady state datasets, characterized by cell type clusters can be analyzed with models that exploit co-expression, prior networks or cell type-specific reference scATAC-seq datasets in combination with sequence motifs to infer directed transcription factor-target relations. **B)** Topological comparison between personalized networks of groups of individuals can reveal coordinated

differences, for instance the change of connectivity in densely connected modules, change of connectivity of hub genes or changes of module membership of individual genes. These differences may help to explain for example the interindividual variation in drug response.

The incorporation of dynamic information extracted from time series or pseudotime[67,68] is another promising avenue to further improve single-cell GRN reconstruction. However, not all datasets are equally well suited to identify temporal trajectories. For example, PBMCs are usually in steady state, and only after pathogen stimulation such trajectories would appear.

Summarized, the ideal GRN reconstruction tool can efficiently manage large amounts of single-cell data, incorporate prior information, model non-linear relationships and take dynamic information into account. Early benchmark studies, performed for a limited number of methods on rather small datasets[45] or on simulated data[69] show that current tools usually only work well in specific situations. As such, there is a clear need for the development of all-round tools that work well in every situation.

**sc-eQTLGen: a federated single-cell eQTL meta-analysis consortium**

Combining data of numerous groups increases the resolution and power by which downstream analyses, such as eQTL identification and personalized GRN reconstruction, can be performed. Ideally, all scRNA-seq datasets should be jointly analyzed at one centralized location. This is particularly helpful to align each group's approaches for preprocessing, quality control (QC) and cell type classification. However, it also eases for instance benchmarking different statistical and computational methods. While this concept of 'bringing the data to the algorithm' is preferred from an analytical perspective, it is usually very difficult to do so when handling privacy-sensitive scRNA-seq and genotype data from human individuals[5,70].

To overcome this, a federated approach could be used instead, which has the aim of 'bringing the algorithm to the data': each participating cohort will run the analyses themselves (adhering to predefined criteria for preprocessing and QC), and will only share summary statistics that are not privacy-sensitive. Finally, one site takes responsibility for performing the overall meta-analysis using these provided summary statistics. For genome-wide association studies this is a common strategy[71,72], and for eQTL studies this procedure has been shown to be effective as well[6,34]. In the following sections we will expand on all steps that have to be taken and what considerations should be made when conducting such a federated approach for single-cell population genetics studies (Figure 4).

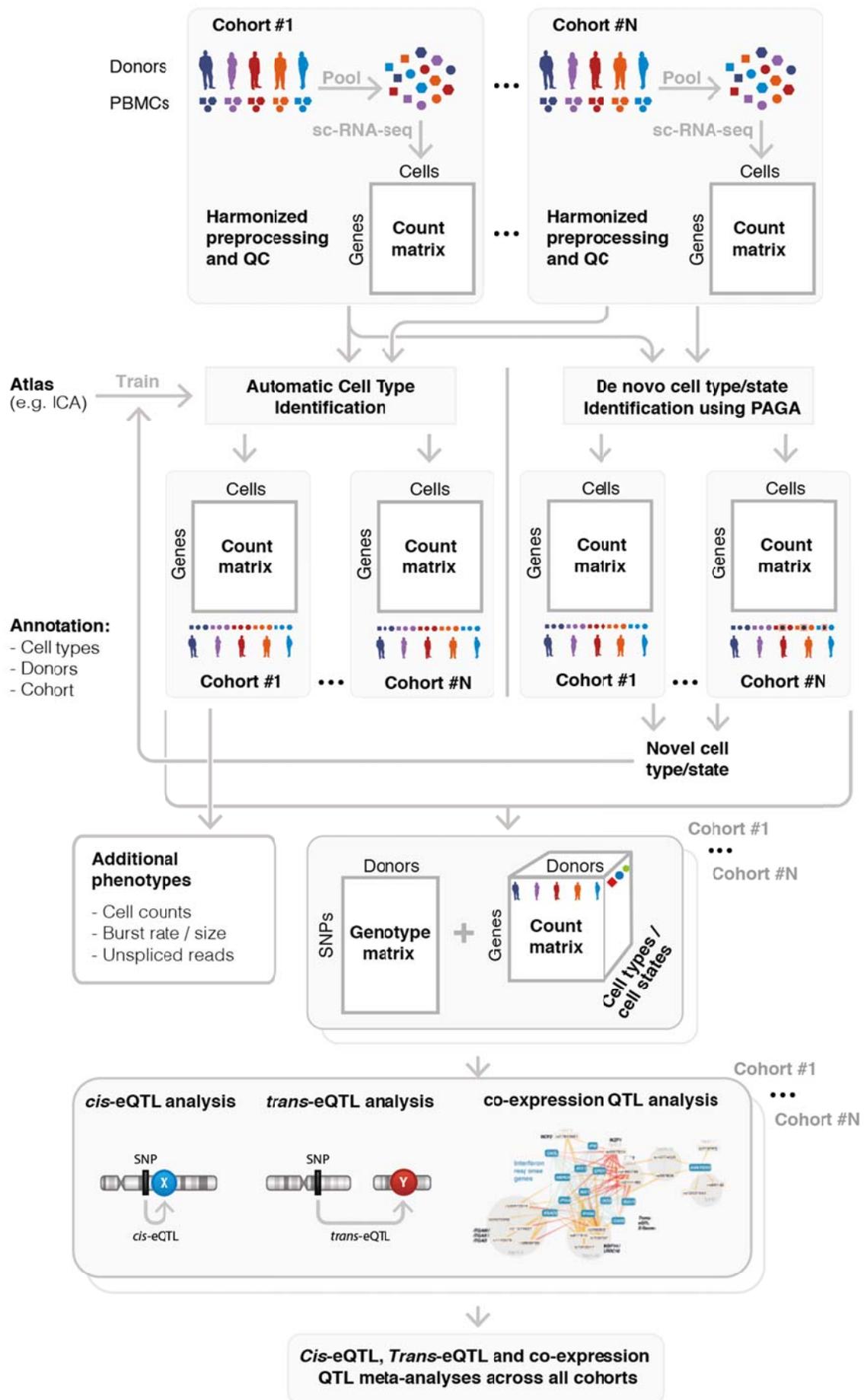

**Figure 4. Overview of the sc-eQTLGen proposed federated approach.** sc-eQTLGen aims to identify the downstream consequences and upstream interactors of gene expression regulation. To increase the resolution and power of this analysis, datasets of multiple cohorts need to be combined while taking privacy issues into account. This will be done using a federated approach in which we will first harmonize all preprocessing and quality control (QC) steps across cohorts. Subsequently, shared gene expression matrices will be normalized and cell types will be classified based on a trained reference dataset (e.g. Immune Cell Atlas (ICA)). Any cells that cannot be classified using this trained classifier, representing new cell types or previously unknown cell states, can then be manually annotated based on marker genes, and then be used to further train the classifier. Each cohort will then separately perform a *cis*- and *trans*-eQTL and co-expression QTL analysis using their genotype and expression matrix, while using appropriate statistical models to account for effects such as gender, population structure and family-relatedness that can alter the genotype-expression relationship in a cohort-specific manner. The summary statistics will be shared and analyzed in one centralized place. Finally, these results will be used for reconstruction of personalized and context-specific gene regulatory networks.

*Preprocessing, quality control*

The first challenge of federated analyses is the need to have a standardized protocol on how each group should perform their analyses. While such a protocol helps to ensure reproducibility of the data analysis, it requires that all methods and tools used have been rigorously tested before. For scRNA-seq data such protocols are still under development, while in other fields such as that of genome-wide association studies, standardized protocols have been available for years.

Several initiatives are now being undertaken to define best practices in the scRNA-seq field[73]. For example, Tian et al. have compared 3,913 combinations of different scRNA-seq data analysis pipelines to define best practices in the field[74]. Such initiatives could provide the basis for defining the optimal preprocessing, QC and cell type classification steps for our consortium. Additionally, in population-based scRNA-seq studies special attention is required to account for ethnic variation and population stratification (Box 1)[75,76]. In the event of presence of relatedness in a given cohort, a genetic relatedness matrix will be included in a mixed model to account for the effect, such as in [76,77]. Adjustments of cohort-level genetic differences will be made in the framework of meta-analysis using summary statistics of the individual cohorts. Once all protocols are established, we can harmonize the preprocessing steps across all groups in the consortium, such as the genome build to use, alignment tool and sample demultiplexing strategy. Due to the cohort-specific characteristics of each dataset, the QC steps cannot be harmonized to the same extent as the preprocessing. Nevertheless, the parameters used for QC can be coordinated across all groups, such as the cutoffs for number of detected genes per cell and mitochondrial fraction. Both the preprocessing and the QC do not require exchanges of data and can be performed independently.

*Cell type classification*

To facilitate the eQTL meta-analysis, we need to ensure that the cell type annotations are consistent across the different cohorts. To ensure reproducibility of annotations across the different cohorts, we will employ a classification scheme to identify canonical cell types in each cohort separately. Performing cell type labeling using classification models does not only increase the reproducibility, but also constitutes a privacy-safe way of annotating cell types that does not require the sharing of raw or processed gene expression data.

Reference datasets with labeled cells, such as those available from the Immune Cell Atlas (http://immunecellatlas.net/) will be used to train a classifier for automatic cell type classification in each cohort. Our recent comparisons of single cell classification methods showed that simple linear models can yield good results[78,79]. Despite the wide availability of reference datasets, we expect that some cohorts will contain novel unknown cell types or states that cannot be identified using the trained classifier. For this, we will use a classification scheme with a rejection option that can flag unknown cells whenever the confidence in cell type assignment is low[78]. The rejected cells can then be manually annotated based on marker gene expression.

To capitalize on the large number of cells and individuals to be profiled in each cohort, an unsupervised clustering approach will be used to analyze the count matrix of each cohort, in parallel to the supervised approach described earlier. This unsupervised approach will serve two purposes: (1) it will help annotate unassigned cells by the classifier, and (2) it will allow refining the resolution at which cells are annotated. Varying levels of granularity of the clustering may reveal cell types, as well as particular cell states or subtypes. This level of granularity required to separate particular cell states is not known a priori. Therefore, novel unbiased approaches such as partition based graph abstraction[80] or metacells, i.e. disjoint, homogenous and highly compact groups of cells that each exhibit only sampling variance[81], provide a framework to reconcile discrete states at different levels of granularity with continuous cell states. These novel annotations can feed back into an iterative online learning approach of supervised classification models, where we could refine cell type prediction models on the available datasets. Once new datasets become available within the consortium these can be annotated based on current models and updated labels can be used in the next round of training. An important consideration here is to preserve the hierarchy of cell annotations, so that if new annotations are added to the classifier, they are subclasses of existing classes. In this way, any downstream analysis based on older annotations remains valid at the older level of granularity. This would yield a coherent approach of labelling over time as the dataset grows. For inference of continuous cell states, we require data integration across multiple centers, as this would ensure the usage of a similar pseudotime scale between individuals. Currently, ordering cells along pseudotime is challenging and best practices are being evaluated[74,82].

Ultimately, integrating all expression data in a privacy-preserving manner, i.e. as gene expression matrices, will produce a dataset with unprecedented numbers of cells. Such a dataset allows discovery of novel rare cell types or states using clustering approaches as described above. This valuable dataset will then be shared with the community through platforms like the HCA data portal.

*eQTL and co-expression QTL analysis*

After cell type assignment, annotated gene expression matrices can be returned to each of the cohorts. Each cohort will then map genome-wide cell type-specific *cis-* and *trans-*eQTLs by combining the cell type-specific gene expression matrices with the privacy-sensitive bulk-assessed genotype information using appropriate statistical models. The resulting summary-statistics can then be safely shared without privacy-issues.

One major problem with federated eQTL analyses is that the amount of summary statistics that need to be shared is substantial. For instance, when assuming there are 10 cohorts and for each of these cohorts cells have been assigned to 10 major cell types, a genome-wide *trans-*eQTL analysis (testing the effect of 10,000,000 common SNPs on 20,000 protein coding genes for each of the 10 cell types), where only the correlation for a SNP-gene combination is stored as a 64 bit double value, would require each cohort to exchange 10,000,0000 x 20,000 x 10 x 8 bytes = 146 terabytes of data. To overcome this problem, several frameworks have recently been proposed that take advantage of the fact that these summary statistics matrices reflect the product of a normalized genotype matrix and a normalized gene expression matrix. For instance, the HASE framework[83] recodes genotype and phenotype (i.e. gene expression) data, along with a covariate matrix, in such a way that privacy is ensured and only those matrices, making up only a few gigabytes of data, need to be exchanged.

While protocols exist that explain how eQTL data needs to be processed, harmonized and QCed to perform a federated eQTL analysis (e.g. eQTLGen used the eQTLMappingPipeline[6]), not all steps can be completed immediately: for instance, to identify effects of polygenic risk scores on gene expression levels (ePRS), gene expression data first needs to be corrected for *cis-*eQTL effects[6]. Therefore, the full *cis-*eQTL meta-analysis has to precede calculations of ePRSs. Such iterations take considerable time and are also inconvenient, since it requires a lot of coordination with each of the participating cohorts. For sc-eQTLGen we will first conduct a federated, cell-type specific *cis-* and *trans-*eQTL analysis. After this is completed, we will proceed with a co-expression QTL (co-eQTL) analysis. This analysis will be limited to a predefined set of genes or SNPs, such as the SNP-gene combinations extracted from the identified *cis-* and *trans-*eQTLs or the SNPs located within open chromatin regions that show high interindividual variability, as otherwise trillions of statistical tests have to be conducted (e.g. in [16]: 7,975 variable genes * 7,975 variable genes * 4,027,501 SNPs (MAF ≥ 0.1) = 256,151,580,788,125 tests). Finally, all these results will be combined to reconstruct personalized, cell type-specific GRNs. This multi-step approach will require us to go back and forth between the different cohorts at least twice. Therefore, easy-to-use analysis scripts that can be run efficiently on different high-performance cluster infrastructures are essential to limit the amount of hands-on time.

*Gene regulatory network reconstruction*

Finally, the scRNA-seq data will be used to reconstruct GRNs. Two strategies will be explored in the context of sc-eQTLGen. The first approach makes use of the large number of bulk RNA-seq datasets for specific cell types that are available in public RNA-seq repositories[84,85]. Using this publicly available bulk RNA-seq data, reference co-expression networks will be constructed using cell type-specific data. Subsequently, scRNA-seq data will be used to implement directionality and specify the connections in the network that are affected by specific contexts[40]. The second approach will directly use scRNA-seq data to build cell type-specific GRNs, thereby enabling to immediately take the context-specificity into account. However, the number of genes that can confidently be taken into account by this second approach may be lower due to the sparsity of scRNA-seq data. For both strategies we will determine the additional benefit of implementing prior information, extracted from either bulk or single-cell data[46,47], and gene expression imputation[46,48]. We expect that the optimal strategy will depend on the amount of available bulk data and prior information that is available for a particular cell type.

Once reconstructed, these GRNs can be used to determine how for example, genetic differences or disease status change the architecture of the network. These networks consist of nodes, representing genes, that are connected through edges, representing the relationship between genes. The context-specific changes in the network can be identified on different levels, such as on the level of individual edges or nodes, topological properties of individual nodes, such as their connectivity (degree) or module membership[86], subnetwork properties, such as the existence and size of modules, or global topological properties, such as degree distribution (Figure 3). Comparing topological features such as node degree to genotypes may identify polymorphisms altering the function of master regulators (highly connected 'hub' genes). Interestingly, implementation of network information was shown to be complementary to the identification of eQTLs, as it identified novel SNPs under genetic control that could not be identified in the single- or multi-tissue eQTL analysis of GTEx[87]. This clearly shows the complementarity of both eQTL and network-based analyses for understanding the impact of genetic variation.

Ultimately, CRISPR perturbations will be coupled to scRNA-seq to validate or improve reconstructed GRNs. To optimize the number of perturbations required for extracting the most useful information from such experiments, an iterative approach will be taken that feeds back the experimental data to the GRN. This approach will make use of active machine learning to select those perturbations that are required to further improve the model[88,89]. These well validated, personalized and context-specific GRNs will provide us with a better understanding of disease and can be the starting point of applying this knowledge for precision medicine in the future.

**Future clinical implications**

The goal of the sc-eQTLGen consortium is to identify how genetic and environmental factors affect gene expression and its regulation in the context of both health and disease. This information will allow us to reveal new targets for disease prevention and treatment (Figure 5). For example, a novel subset of tissue-resident memory T cells has recently been identified in the setting of asthma using scRNA-seq[90]. This study also showed that mostly T helper 2 cells are dominating the cell-cell interactions in the asthmatic airway wall, whereas in healthy controls mostly epithelial and mesenchymal cell types are communicating with each other. Integration of the gene expression of this asthma-associated cell type with asthma-associated genetic risk variants would further increase our understanding of the disease and such knowledge would greatly accelerate the development of personalized/precision treatments in the future. It is this information about how genes interact differently between individuals as a function of their genetic predisposition that will be obtained through the results of our consortium (Figure 5). One of the major benefits of such personalized treatments is in prescribing the correct drug based on the individual (mechanism that underlies) susceptibility to disease. Currently only between 4% and 25% of the people respond to commonly prescribed drugs[91], showing the need to better predict drug responsiveness and thereby avoid unnecessary exposure to side-effects.

This high interindividual variability in drug response is a consequence of genetic and environmental exposure differences between individuals, which can result in differences in drug metabolism, absorption and excretion (pharmacodynamics)[92]. For example, a variant in the *CYP2C19* gene changes the response to the anti-blood clotting drug clopidogrel. The *CYP2C19* gene encodes for an enzyme in the bioactivation of the drug. *CYP2C19* poor metabolizers were shown to exhibit higher cardiovascular event rates after acute coronary syndrome, or percutaneous coronary intervention, as compared to patients with normal *CYP2C19* function[93].

While previous efforts have mainly focused on pharmacodynamic variation, recent single-cell analyses have revealed that gene-gene interactions can also be changed by genetic[16] and environmental variation[10]. For example, two closely related SNPs (linkage disequilibrium $R^2$ = 0.92) affected both gene-gene interactions (*RPS26* and *RPL21*)[16] and gene-environment interactions (*RPS26* and the respiratory status of the cell)[10]. This shows that gene regulatory network changes may underlie part of the interindividual variation in drug responsiveness. However, such effects have never been studied in detail before and the extent to which such interactions affect drug responsiveness are unknown. The sc-eQTLGen consortium is able to study both how gene-gene interactions and gene-environment interactions are affected by genetic variation, giving insight into where and when they occur. Importantly, the applied methodologies will be easily transferable to single-cell data that is collected in other cell types and disease context through other large-scale efforts[19] (https://lifetime-fetflagship.eu). Moreover, several partners within our consortium have generated scRNA-seq data in cohorts with extensive information on individuals' health records and drug usage (e.g. Lifelines Deep cohort[94] and the OneK1K cohort). With such information, we will be able to validate the link between changes in the gene regulatory network and the drug responsiveness of an individual. This allows us to determine the predictive value of gene networks for determining responsiveness of specific drugs and the applicability of such networks in precision medicine. As such, the sc-eQTLGen consortium will not only increase our basic knowledge about the contribution of genetics in gene expression and its regulation, but will also be a valuable resource for drug target identification and validation.

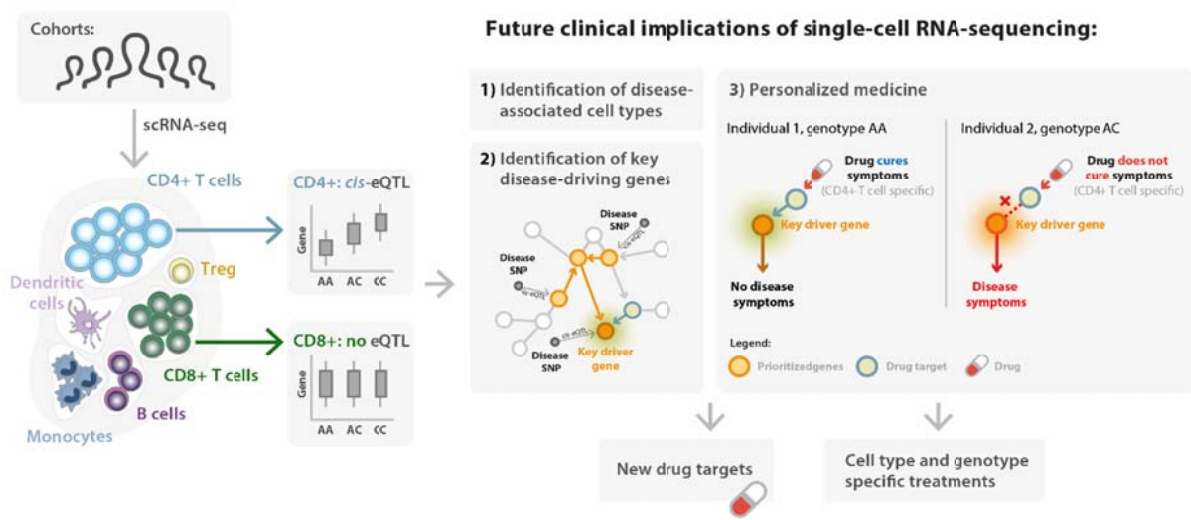

**Figure 5. Future clinical implications of the single-cell eQTLGen Consortium.** The efforts of sc-eQTLGen will lead to the (**1**) identification of disease-associated cell types and (**2**) key disease-driving genes, which together will aid (**3**) the implementation of personalized medicine and the development of new therapeutics that take all this information into account (cell type- and genotype-specific treatments).

**Box 1: Guidelines for creating a population-based single-cell cohort**

Population-based single-cell datasets have proven to be a valuable addition to bulk-based datasets for studying the effects of genetic variation on gene expression and its regulation[16, 23]. In comparison to 'standard' single-cell datasets, generating such population-based single-cell datasets require some additional aspects to be taken into account.

First of all, the genetic information that is available for each of the individuals in such cohorts can be used to demultiplex pools of multiple individuals within the same sample. This approach allows to properly randomize experiments, while also significantly reducing cost and confounding effects[23, 95]. This genetic information can either be efficiently generated using genotype arrays[96] in combination with imputation-based approaches[97], or extracted from the scRNA-seq data itself[95, 98]. The basic principle behind genetic multiplexing is that enough transcripts harboring SNPs are expressed and detected in each single cell such that cells can be accurately assigned to the donor of origin. Furthermore, as the number of multiplexed individuals increases, the probability that a droplet harbors multiple cells from different individuals increases, thus allowing the detection of multiplets using genetic information. This enables the overloading of cells into standard droplet-based workflows and overall reduction of cost per cell up to about 10-fold (https://satijalab.org/costpercell). As the cost of sequencing and the background multiplet rate reduce, the benefits of multiplexing increase. We anticipate that future workflows will allow for even higher throughput.

Secondly, accounting for ethnicity variation and population stratification will be required when single-cell data of diverse populations are being analyzed. It is known that a different genetic architecture exists between different populations. Nevertheless, practical considerations have limited the majority of eQTL studies to cohorts of European origin. As an undesirable consequence of this bias in population representation, certain variants may not have been detected before[99] or the effect sizes and associated polygenic risk scores based on the European population may not be translatable to other populations[100, 101]. Therefore, inclusion of datasets from different ethnic populations will help reduce long-standing disparities in genetic studies and has many analytical advantages[102, 103]. For example, the increased genotype frequency diversity will enhance the range over which gene expression varies, and thereby, will further increase statistical power. To implement multi-population sc-eQTL analysis, several challenges have to be addressed. Handling data from populations with different levels of population genetic properties such as LD structure, relatedness and multiple genetic origins that result in the presence of genetic covariance remains important and requires appropriate adjustments to avoid spurious signal and to manage the bias in estimating genetic *cis*- and *trans*-effects[76, 104]. This is particularly important when differences in cohort-specific genetic characteristics are enhanced such as when family-based and unrelated cohorts or cohorts of different ancestries are analyzed. Failing to account for these effects affects the accuracy of mapping and results in false positives.

Finally, studying genetic variation at the single-cell level adds some extra requirements for the number of cells per individual and the number of individuals to be included in the study. The number of cells per individual will mainly define for which cell types in a heterogeneous sample such as PBMCs eQTL and co-eQTL analyses can be performed. In contrast, the number of individuals will mainly define the number of genetic variants for which effects on gene expression can be confidently assessed. A recent analysis showed that, with a fixed budget, the optimal power for detecting cell type-specific eQTLs is obtained when the number of reads is spread across many individuals[105]. Even though a lower sequencing depth per cell results in a lower accuracy of estimating cell type-specific gene expression levels, many more individuals and cells per individual can be included for the same budget. As a result, the optimal experimental design with a fixed budget provides up to three times more power than a design based on the recommended sequencing depth of 50,000 reads per cell (for 10X Genomics scRNA-seq). In contrast, for co-eQTL analysis there is a different trade-off between sequencing depth, number of individuals and number of reads per cell; while for eQTL analysis gene expression levels among cells of the same cell type can be averaged, for co-eQTL analysis you cannot as this would prohibit you from calculating a gene-gene correlation per individual. Therefore, for co-eQTLs the sequencing depth will be a major limiting factor that determines the number of genes for which you can confidently calculate gene-gene correlations. Altogether, depending on the goal of your study, the optimal balance between sequencing depth and number of individuals and cells per individual will be different.

**Competing interest**

The authors declare no competing interest.